\documentstyle[seceq,epsf,wrapfig,twoside]{ptptex}
\notypesetlogo  
\title{The Present Status on $\sigma$ and $\kappa$ Meson Properties 
}
\author{%
Muneyuki {\sc Ishida}
}
\inst{%
Department of Physics, Tokyo Institute of Technology\\
Tokyo 152-8551, Japan
}
\recdate{%
\today
}
\abst{%
The recent experimental data of both $\pi\pi /K\pi$ scattering and production processes, 
suggesting the existence of scalar $\sigma$ and $\kappa$ mesons,
are reviewed. 
In many $\pi\pi /K\pi$ production processes the direct effects of their productions
are observed, while they are, because of chiral symmetry, hidden in scattering processes, 
and now $\sigma$(500$\sim$600) and 
$\kappa$(800$\sim$900) are considered to be confirmed experimentally. 
The recent criticism on our method of analyses,
which is based on the long believed prejudice of 
universal $\pi\pi /K\pi$ phase through scattering and production
amplitudes, is explained not to be valid.
}

\begin{document}
\maketitle

\setcounter{tocdepth}{4}

\section{Introduction}

The iso-singlet scalar $\sigma$ meson was introduced theoretically in 
the linear $\sigma$ model\cite{Sch}, and its existence was first suggested 
in one-boson-exchange potential model\cite{OBEP} of nuclear forces. 
The importance of $\sigma$ was stressed\cite{Sca,HK} 
in relation with dynamical chiral symmetry breaking of QCD. 
However, its existence had been neglected 
phenomenologically for many years, being based on the negative results\cite{pen} 
of $\pi\pi$ scattering phase shift analyses.

Recently the $\pi\pi$ phase shift\cite{CM} was reanalyzed by many groups\cite{PS}
including ours\cite{555} and the existence of light $\sigma (450$-$600)$ 
was strongly suggested. 
The result of the previous analysis with no $\sigma$ existence 
was pointed out\cite{frascati} to be not correct, since in this analysis\cite{pen}
there is no consideration on the cancellation mechanism between 
$\sigma$ amplitude and 
non-resonant $\pi\pi$ amplitude, which is guaranteed by chiral symmetry.  

On the other hand, 
it is remarkable that, in contrast with the spectra of $\pi\pi$ scattering, 
the clear peak structure has been observed in mass region of 
$m_{\pi\pi}\sim 500$ MeV
in the various $\pi\pi$ production processes, such as 
$J/\psi \rightarrow \omega\pi\pi$,\cite{DM2,had97Jpsi,WuNing}
 $p\bar p\rightarrow 3\pi^0$,\cite{CB,ppbar}
$D^+\rightarrow \pi^+\pi^-\pi^-$\cite{E791} and 
$\tau^-\rightarrow \pi^-\pi^0\pi^0 \nu_\tau$,\cite{tau} and this structure 
is shown to be well reproduced by the Breit-Wigner amplitude of $\sigma$ meson.

Thus, presently firm evidences\cite{YITP} of $\sigma$ seem to be accumulated, 
and  
the column of $\sigma$ in particle data group table is corrected as 
``$f_0(600)$ or $\sigma$" in the newest'02\cite{PDG} edition 
in place of $f_0(400$--1200) or $\sigma$ in the '96--'00 editions.
There are now hot controversies on the existence of $I=1/2$ scalar 
$\kappa$ meson, to be assigned as a member of $\sigma$ nonet.
Reanalyses\cite{kappa,Kpi} of $K\pi$ scattering phase shift\cite{LASS} 
suggest existence of the $\kappa (900)$, while no $\kappa$ is insisted in ref. \citen{penkappa}.
The existence of $\kappa$ is again suggested strongly
in $K\pi$ production process of $D^+\rightarrow K^-\pi^+\pi^+$,\cite{E791kappa}
similarly to the case of $\sigma$.

In the analyses of the $\pi\pi$ (or $K\pi$) production processes mentioned above,
the amplitudes are parametrized by a coherent sum of the Breit-Wigner amplitudes 
including $\sigma$ (or $\kappa$) and the non-resonant $\pi\pi$ (or $K\pi$) production amplitude. 
This parametrization method is called VMW method. 

Recently there have been raised some criticisms\cite{MO1,bug} on the VMW method,
especially in relation to the consistency with $\pi\pi$ 
(or $K\pi$) scattering phase shift. However, as will be clarified in \S 3,
we emphasize that the production amplitude is, in principle, independent from the scattering
amplitude and that 
the analyses of production processes 
should be done independently of scattering processes.
Also it is shown that our method of analyses, VMW method, is consistent\cite{PLB2} 
with all the constraints from unitarity and chiral symmetry. 

In this paper we review both of the $\pi\pi$($K\pi$) scattering and 
production processes
relevant to $\sigma$ meson ($\kappa$ meson), 
and  report the present status of $\sigma$ and $\kappa$ meson properties.
The phase motion of the production amplitude is also examined, and 
the above criticisms on VMW method will be clarified not to be correct.

\section{Experimental Evidences for $\sigma$ and $\kappa$}

\begin{wrapfigure}{r}{4.5cm}
  \epsfysize=6.5cm
   \centerline{\epsffile{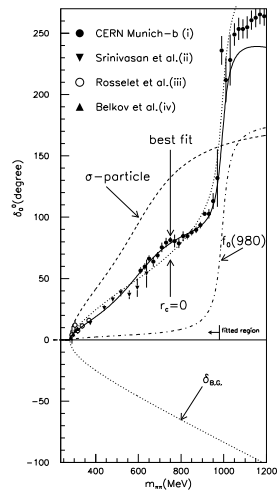}}
  \caption{The fit to $I=0$ $S$ wave $\pi\pi$ scattering phase shift including
    $\delta_{BG}$ of hard-core type. 
    The result is compared with the conventional analysis 
    with no $\delta_{BG}^{Non.Res.}$ ($r_c=0$). 
    The greatly improved $\chi^2$
    strongly suggests existence of $\sigma (600)$.
 }
  \label{figpipi}
\end{wrapfigure}
({\it $\pi\pi$ scattering})\ \ \ \ 
We first review our reanalysis\cite{555} of 
$\pi\pi$ scattering phase shift $\delta$ obtained by CERN-Munich.\cite{CM}
The $\delta$ of $I=0$ $S$ wave amplitude, $\delta_S^0$,
is fitted by Interfering Amplitude method, 
where the total $\delta_S^0$ below $m_{\pi\pi}\simeq 1$GeV
is represented by the sum of the component phase shifts,
\begin{eqnarray}
\delta_S^0 &=& \delta_\sigma + \delta_{BG} + \delta_{f_0} .
\label{b1}
\end{eqnarray}
The $\delta_\sigma$ is from $\sigma$ Breit-Wigner amplitude 
and $\delta_{f_0}$ is from $f_0(980)$ Breit-Wigner 

amplitude with narrow width. The $\delta_{BG}$ is from 
non-resonant repulsive $\pi\pi$ amplitude, and is taken 
phenomenologically of hard-core type, $\delta_{BG}=-p_1 r_c$
($p_1=\sqrt{s/4-m_\pi^2}$ being the CM momentum of $\pi$).

The experimental $\delta_S^0$ passes through 90$^\circ$ at 
$\sqrt s(=m_{\pi\pi})\sim 900$MeV. This is explained by the 
cancellation between attractive $\delta_\sigma$ and repulsive 
$\delta_{BG}$. 
The result of the fit is given in Fig. \ref{figpipi}.
The mass and width of $\sigma$ is obtained as $m_\sigma =585\pm 20$MeV and
$\Gamma_\sigma =385\pm 70$MeV.

Note that the above cancellation is shown\cite{frascati} to come from chiral symmetry
in the linear $\sigma$ model (L$\sigma$M): 
The $\pi\pi$ scattering $A(s,t,u)$ amplitude in L$\sigma$M is given by 
\begin{eqnarray}
A(s,t,u) &=& \frac{(-2g_{\sigma\pi\pi})^2}{m_\sigma^2-s} - 2 \lambda 
 = \frac{s-m_\pi^2}{f_\pi^2} + \frac{1}{f_\pi^2} 
\frac{(s-m_\pi^2)^2}{m_\sigma^2-s},\ \ \ \ \ 
\label{b2}
\end{eqnarray}
as a sum of the $\sigma$ amplitude $A_\sigma$, which is strongly attractive, 
and of the non-resonant $\pi\pi$ amplitude $A_{\pi\pi}$ due to the $\lambda\phi^4$ 
interaction, which is stongly repulsive.
They cancel with each other following the relation of L$\sigma$M,
 $g_{\sigma\pi\pi}=f_\pi\lambda=(m_\sigma^2-m_\pi^2)/(2f_\pi)$,
and the small ${\cal O}(p^2)$ Tomozawa-Weinberg (TW) amplitude and 
its correction are left.
The $A_\sigma (A_{\pi\pi})$ corresponds to $\delta_\sigma$($\delta_{BG}$).

\begin{wrapfigure}{r}{6cm}
  \epsfysize=3.cm
  \centerline{\epsffile{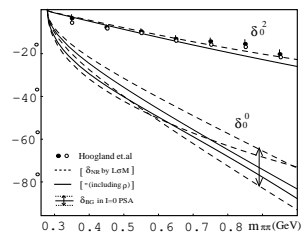}}
  \caption{The phase $\delta_{NR}$ of $I=0,2$ calculated 
from non-resonant part of the amplitudes
by L$\sigma$M (thick dashed line) and those obtained by L$\sigma$M including $\rho$ meson
contribution (thick solid line): 
The $N/D$ unitarization is used for tree level amplitude. 
(See ref. \cite{frascati} for detail.)
The phenomenological $\delta_{BG}=-p_1r_c (r_c=0.60\pm0.07{\rm fm})$\cite{555} in $I=0$ 
is shown by thin lines with errors. 
The experimental $\delta_0^2$\cite{hoog} is also shown.
 }
  \label{figBG}
\end{wrapfigure}
Actually, as shown in Fig.~\ref{figBG}, 
our theoretical curves for $\delta_{NR}^0$ and $\delta_{NR}^2$, obtained by unitarizing the
respective Born amplitudes, given by $A_{\pi\pi}$, $A(t,s,u)$ and $A(u,t,s)$,
in L$\sigma$M are consistent with our
$\delta_{BG}$ of hard core type in our phase shift analysis,\cite{555}
and with 
experimental $\delta_0^2$,\cite{hoog} respectively. 

Thus, it is shown that the $\sigma$ Breit-Wigner amplitude with
non-derivative (${\cal O}(p^0)$) $\pi\pi$-coupling
requires at the same time
the strong (${\cal O}(p^0)$) repulsive $\pi\pi$ interaction to obtain
the small ${\cal O}(p^2)$ TW amplitude, satisfying chiral symmetry. 
This is the origin of
$\delta_{BG}$ in our phase shift analysis.

There was an argument\cite{MO2} that a broad resonance with mass 1GeV,
denoted as $f_0(1000)$\cite{CM} or $\epsilon (900)$,\cite{pen}
instead of light $\sigma$, exists.
We already investigated this possiblity.
The fit with $r_c=0$ corresponds to the conventional analyses 
without the repulsive $\delta_{BG}^{Non.Res.}$.
Actually the pole position of ``$\sigma$" of this fit, $\sqrt s=970-i320$,
is very close to $\sqrt s=1046-i250(910-i350)$MeV of $f_0(1000)$($\epsilon (900)$).
The resulting $\chi^2$ is $\chi^2/N_F=163.4/31$ (See Fig.~\ref{figpipi}).
When we take into account the cancellation mechanism of chiral symmetry
by including $\delta_{BG}^{Non.Res.}$, as was done in the present analysis,
$\chi^2/N_F=23.6/30$ is obtained.
The greatly improved $\chi^2$
strongly suggests the $\sigma (600)$, rather than $f_0(1000)$.

({\it $K\pi$ scattering})\ \ \ \ 
The similar cancellation is also expected to occur in $K\pi$ scattering since
$K$ has also a property of Nambu-Goldstone boson. 
The $I=1/2$ $S$ wave phase shift
$\delta_{S}^{1/2}$ is parametrized by introducing the
$\kappa$ Breit-Wigner phase shift $\delta_\kappa$ and 
its compensating repulsive
non-resonant $K\pi$ phase shift $\delta_{BG}^{Non.Res}$ as
$\delta_{S}^{1/2}=\delta_\kappa + \delta_{BG}^{Non.Res}+\delta_{K_0^*(1430)}$.
The fit to 
$\delta_{S}^{1/2}$ below 1.6 GeV by LASS\cite{LASS} gives the 
mass and width of $\kappa$ meson\cite{Kpi} as 
$m_\kappa =905\stackrel{+65}{\scriptstyle -30}$ MeV
and $\Gamma_\kappa =545\stackrel{+235}{\scriptstyle -110}$ MeV.

({\it $J/\psi\rightarrow\omega\pi\pi$})\cite{DM2}\ \ \ \ 
Following VMW method,
the production amplitude is represented by a sum of Breit-Wigner amplitudes 
of $\sigma$ and $f_2(1275)$ and the non-resonant background.\cite{had97Jpsi} 
The $m_{\pi\pi}$ mass spectra clearly shows a peak structure, which is explained
by a Breit-Wigner amplitude of $\sigma$ with 
$(m_\sigma ,\Gamma_\sigma )=(482\pm 3,325\pm 10)$MeV.
This is quite in contrast with the situation in the $m_{\pi\pi}$ spectra 
of $\pi\pi$ scattering,
where the no direct $\sigma$ peak is observed because of the cancellation mechanism
of chiral symmetry. The similar result is obtained by BES\cite{WuNing} with 
$(m_\sigma ,\Gamma_\sigma )=(390\stackrel{+60}{\scriptstyle -36},
282\stackrel{+77}{\scriptstyle -50})$ MeV.

({\it $\Upsilon "\rightarrow \Upsilon \pi\pi$})\ \ \ \ 
The $m_{\pi\pi}$ spectra of $\Upsilon (2S)\rightarrow \Upsilon (1S)\pi\pi$, 
$\Upsilon (3S)\rightarrow \Upsilon (1S)\pi\pi$, 
$\Upsilon (3S)\rightarrow \Upsilon (2S)\pi\pi$, 
$\psi (2S)\rightarrow J/\psi\pi\pi$, $J/\psi \rightarrow \phi \pi\pi$ and $\phi K\bar K$
are commonly fitted\cite{PLB1} by the amplitude by VMW method, 
which is a coherent sum of 
$\sigma$ Breit Wigner amplitude ${\cal F}_\sigma$ and 
non-resonant $\pi\pi$ amplitude ${\cal F}_{2\pi}$.
The $(m_\sigma ,\Gamma_\sigma )=(526\stackrel{+48}{\scriptstyle -37},
301\stackrel{+145}{\scriptstyle -100})$ is obtained with $\chi^2/N_F=86.5/(150-37)=0.77$.
The double peak structure in $\Upsilon (3S)\rightarrow \Upsilon (1S)\pi\pi$
is explained by the interference between the
${\cal F}_{2\pi}$ with constant phase and the ${\cal F}_{\sigma}$ 
with moving phase.

({\it $p\bar p \rightarrow 3\pi^0$})\ \ \ \ 
The $\pi^0\pi^0$ mass distribution\cite{CB} and the cos$\theta$ distributions 
in $m_{\pi\pi}$
around $K\bar K$ threshold and 1.5 GeV are fitted\cite{ppbar} by the amplitude
of a superposition of $\pi^0 R$ amplitudes with $R=\sigma , f_0(980,1300,1500)$ and
$f_2(1275,1565)$. The total amplitude symmetrized for three identical $\pi^0$
shows a peak in $m_{\pi\pi}\sim 700$MeV, which is explained by the 
contribution from $\sigma$ Breit Wigner amplitude with 
$(m_\sigma ,\Gamma_\sigma )=(540\stackrel{+36}{\scriptstyle -29},
385\stackrel{+64}{\scriptstyle -80})$. 

({\it $D^+\rightarrow\pi^-\pi^+\pi^+$,
$D^+\rightarrow K^-\pi^+\pi^+$ })\ \ \ \ 
The E791 adopted VMW method in our words, and the Dalitz plot
of $D^+\rightarrow\pi^-\pi^+\pi^+$ is analyzed by
the amplitude\cite{E791} of a coherent sum of the $\sigma$
and the other relevant Breit Wigner amplitudes and 
the non-resonant $3\pi$ amplitude.
The $\pi^+\pi^-$ mass distribution shows a clear peak structure in $m_{\pi\pi} \sim 500$
MeV region, which is explained by $\sigma$ contribution
with $(m_\sigma ,\Gamma_\sigma )=(478\stackrel{+24}{\scriptstyle -23}\pm 17,
324\stackrel{+42}{\scriptstyle -40}\pm 21)$MeV. 
The $\sigma\pi^+$ decay is a main mode in this channel,
and its fraction is $46.3\pm 9.0 \pm 2.1\%$.

Similarly the Dalitz plot of $D^+\rightarrow K^-\pi^+\pi^+$
is analyzed by VMW method and the existence of $\kappa (800)$ is strongly
suggested with  $(m_\kappa ,\Gamma_\kappa )=(797\pm 19 \pm 42,
410\pm 43 \pm 85)$MeV.\cite{E791kappa} 
The $\kappa\pi^+$ decay is a main mode in this channel,
and its fraction is $47.8\pm 12.1 \pm 3.7\%$, which is almost equal to the fraction
of $\sigma\pi^+$ in $D^+\rightarrow\pi^-\pi^+\pi^+$. 
This fact suggests the $\sigma (500)$ and $\kappa (800)$ belong to 
the same nonet.\cite{CloseNils}

({\it $\tau^-\rightarrow \pi^-\pi^0\pi^0\nu_\tau$})\ \ \ \ 
The $3\pi$ mass distribution of $\tau^-\rightarrow \pi^-\pi^0\pi^0\nu_\tau$
by CLEO\cite{tau} is explained mainly by $a_1(1260)$.
It is remarkable that the $\pi^0\pi^0$ mass distributions show clear peak
structures of $\sigma$,  while  $\pi^-\pi^0$ mass distributions show clear peak
of $\rho$. Branching fractions of $\rho\pi^0\nu_\tau$ and $\sigma\pi^-\nu_\tau$
are 60.19$\%$ and $18.76\pm 4.29\%$, respectively.
The $(m_\sigma ,\Gamma_\sigma )$ are (555,540)MeV. 

\section{Method of Analyses of $\pi\pi /K\pi$ Production Processes}

({\it Essence of VMW method})\ \ \ \ 
The analyses of $\pi\pi /K\pi$ production processes 
quoted in the previous section are done following the VMW method.
Here we explain our basic physical picture on strong interactions and the essential
point of this method.

The strong interaction is a residual interaction of QCD among all 
color-neutral bound states of quarks($q$), anti-quarks($\bar q$) and gluons($g$). 
These states are denoted as $\phi_i$, and the strong interaction 
Hamiltonian ${\cal H}_{\rm strong}$ is described by these 
$\phi_i$ fields.
It should be noted that, from the quark physical picture,\cite{ShinMune} 
unstable particles
as well as stable particles, if they are color-singlet bound states, 
should be equally treated as $\phi_i$-fields on the same footing.
\begin{eqnarray}
{\cal H}_{\rm strong} &=& {\cal H}_{\rm strong}  (\phi_i) \nonumber\\
\{\ \phi_i \ \} &=& 
 \{  {\rm color\ singlet\ bound\ states\ of\ }q,\bar q\ {\rm and}\ g   \}    .
\label{c1}
\end{eqnarray}

The time-evolution by ${\cal H}_{\rm strong} (\phi_i)$ describes 
the generalized $S$-matrix. 
Here, it is to be noted that, if  
 ${\cal H}_{\rm strong}$ is hermitian, the unitarity of $S$ matrix is guaranteed.
 
\begin{wrapfigure}{c}{12cm}
  \epsfysize=2.5cm
  \centerline{\epsffile{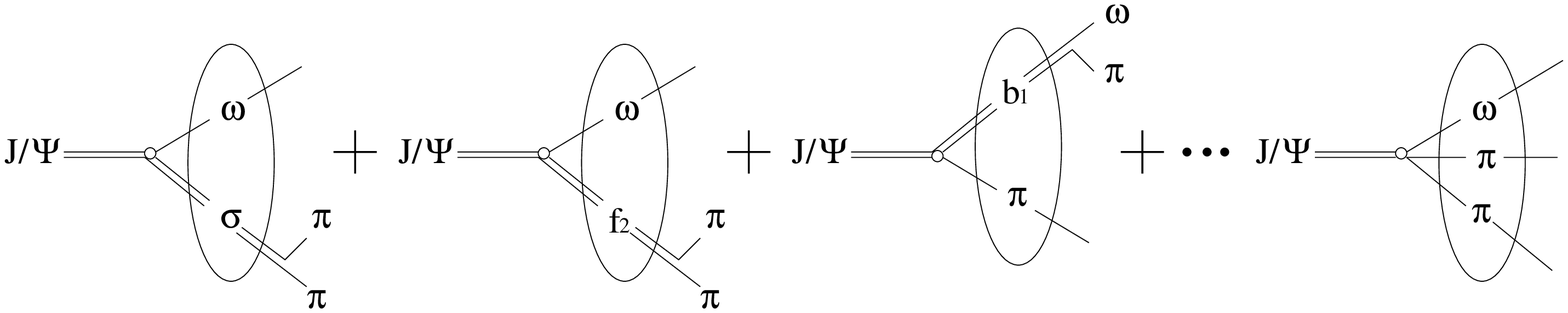}}
  \caption{The final state interactions in $J/\psi\rightarrow\omega\pi\pi$.
The amplitude is a superposition of different $S$ matrix elements, such as
$J/\psi\rightarrow\omega\sigma$,  $J/\psi\rightarrow\omega f_2$,  
$J/\psi\rightarrow b_1 \pi$, $\cdots$,  $J/\psi\rightarrow\omega (\pi\pi)_{Non.Res.}$.
The ellipses represent the final state interactions, 
and the corresponding amplitudes have independent strong phases.   
 }
  \label{figFSI}
\end{wrapfigure}

The bases of generalized $S$-matrix 
are the configuration space of these multi-$\phi_i$ states. 
\begin{eqnarray}
\begin{array}{l} S\ {\rm matrix\ bases}\\ 
 \ \{ {\rm multi-}\phi_i{\rm -states} \} \end{array}
 &=& 
   \left\{  \begin{array}{l}   
  | \omega \pi\pi \rangle , \underline{ | \omega\sigma \rangle , 
                   | \omega f_2 \rangle ,| b_1 \pi \rangle , \cdots , |J/\psi\rangle }, \\
  | N\pi \rangle ,(|N\pi\pi\rangle)_{Non.Res.} ,
                 \underline{ |\Delta\rangle ,|\Delta\pi\rangle ,
         |\Delta\sigma\rangle ,\cdots},\\ 
  \cdots\cdots 
   \end{array}     \right\}  ,\ \ \ \ \ \ \ \ \ \ 
\label{c2}
\end{eqnarray}
where the states relevant for $J/\psi\rightarrow\omega\pi\pi$ decay
and $N\pi$ scattering are respectively shown in 1st and 2nd lines as examples.
The states including unstable particles shown with underlines are
equally treated with non-resonant states $|\omega\pi\pi\rangle$ and $|N\pi\rangle,|N\pi\pi\rangle$.
The relevant $J/\psi \rightarrow \omega\pi\pi$ decay process has the 3-body final state $\omega\pi\pi$.
This process is described by a coherent sum of amplitudes 
for various 2-body decays, 
$J/\psi\rightarrow\omega\sigma$, $J/\psi\rightarrow\omega f_2(1275)$,
$J/\psi\rightarrow b_1(1235)\pi,\cdots$, and for a non-resonant 
3-body($\omega\pi\pi$) decay.
These respective decay amplitudes correspond to different non-diagonal 
elements of the 
generalized $S$-matrix, and have independent coupling strengths.\footnote{
The strengths and phases of respective amplitudes are considered to be determined 
by quark dynamics. However, we treat them independent in phenomenological analyses.
}
The ${\cal H}_{\rm strong}$ induces the various final state interaction, 
reducing to the strong phases of the corresponding amplitudes.
(See Fig.\ref{figFSI}.)  

The remaining problem is how to treat unstable particles, as
there is no established field-theoretical method for this problem. 
In VMW method unstable particles are treated intuitively 
by the replacement of propagator,
\begin{eqnarray}
{\rm Stable\ particle} & & {\rm Unstable\ particle}\nonumber\\
\frac{1}{m_\sigma^2-s-i\epsilon} 
&\stackrel{\rm Strong\ Int.}{\longrightarrow}&
\frac{1}{m_\sigma^2-s-im_\sigma\Gamma_\sigma (s)} ,
\label{c3}
\end{eqnarray}
where we take the case of $\sigma$ as an example.

The effective $\omega\pi\pi$ amplitude is given by a coherent sum of 
all those decay amplitudes,
\begin{eqnarray}
{\mib  {\cal F}_{\mib \omega \mib \pi \mib \pi} }\ \ \  &=& 
   {\cal F}_{\omega\sigma}+ {\cal F}_{\omega f_2}  
   +{\cal F}_{ b_1\pi} +\cdots + {\cal F}_{\omega (\pi\pi)_{Non.Res.}} ,
\label{c4}\\
 {\cal F}_{\omega\sigma}\ \ \  
     &=&  r_\sigma e^{i\theta_\sigma}\  
        \frac{m_\sigma \Gamma_\sigma}{m_\sigma^2-s-im_\sigma\Gamma_\sigma (s)}
       \simeq \ {}_{out}\langle \omega\sigma |\  J/\psi\  \rangle_{in} \nonumber\\
{\cal F}_{\omega f_2}\ \ \  
     &=&  r_{f_2} e^{i\theta_{f_2}} \ 
        \frac{m_{f_2} \Gamma_{f_2} N_{\pi\pi}(s,{\rm cos}\theta)}
              {m_{f_2}^2-s-im_{f_2}\Gamma_{f_2} (s)}
       \simeq \ {}_{out}\langle \omega f_2 |\  J/\psi\  \rangle_{in} \nonumber\\
{\cal F}_{\omega b_1}\ \ \  
     &=&  r_{b_1} e^{i\theta_{b_1}} \ 
        \frac{m_{b_1} \Gamma_{b_1}}
              {m_{b_1}^2-s-im_{b_1}\Gamma_{b_1} (s)}
       \simeq \ {}_{out}\langle \omega b_1 |\  J/\psi\  \rangle_{in} \nonumber\\
     \cdots &&  \nonumber\\
  {\cal F}_{\omega (\pi\pi)}^{Non.Res.} 
     &=&  r_{2\pi}^{N.R.} e^{i\theta_{2\pi}^{N.R.}}
      \ \ \ \ \ \ \ \ \ \ \ \ \ \ \ \ \  
         \simeq \ {}_{out}\langle \omega (2\pi)_{N.R.} |\  J/\psi\  \rangle_{in} \ ,
\label{c5}
\end{eqnarray}
where ${\cal F}_{\omega\sigma}$ corresponds to the generalized $S$ matrix 
element $\ {}_{out}\langle \omega\sigma |\  J/\psi\  \rangle_{in}$, 
$|\ \ \rangle_{in} (\ {}_{out}\langle \ \ |$ denoting $in(out)$-state of 
scattering theory. The $r_\sigma$ represents the corresponding 
coupling strength and the $\theta_\sigma$ comes from 
the $\omega\sigma$ rescattering (the final state interaction).
The extra Breit-Wigner factor comes from the prescription Eq.~(\ref{c3}). 
Similarly, ${\cal F}_{\omega f_2}$, ${\cal F}_{\omega b_1}$ and  
${\cal F}_{\omega (\pi\pi)}^{Non.Res.}$ correspond to the generalized 
$S$ matrix elements,
$\ {}_{out}\langle \omega f_2 |\  J/\psi\  \rangle_{in}$, 
 $\ {}_{out}\langle \omega b_1 |\  J/\psi\  \rangle_{in}$ and  
$\ {}_{out}\langle \omega (2\pi)_{N.R.} |\  J/\psi\  \rangle_{in}$, 
respectively, and they have mutually-independent couplings, 
$r_{f_2}$, $r_{b_1}$ and $r_{2\pi}^{N.R.}$, 
and strong phases, $\theta_{f_2}$, $\theta_{b_1}$ 
and $\theta_{2\pi}^{N.R.}$. $N_{\pi\pi}$ is angular function
of $f_2\rightarrow \pi\pi$ $D$ wave decay.

({\it Relation between scattering and production amplitudes, and chiral constraint})\ \ \ \ 
In $\pi\pi$ scattering the derivative-coupling 
property of Nambu-Goldstone $\pi$-meson requires 
the suppression of the amplitude
${\cal T}_{\pi\pi}$ near threshold,
\begin{eqnarray}
{\cal T}_{\pi\pi} &\sim& -p_{\pi 1}\cdot p_{\pi 2} \rightarrow  m_\pi^2 \sim 0 \ \ \ {\rm at} \ \ \ 
s\rightarrow 4m_\pi^2.
\label{c6}
\end{eqnarray}
This chiral constraint requires, as was explained in \S 2, the
strong cancellation between the $\sigma$ amplitude ${\cal T}_\sigma$ and the
non-resonant $\pi\pi$ amplitude ${\cal T}_{2\pi}$, which 
means the strong constraints, $r_{2\pi}\simeq -r_\sigma$, 
$\theta_\sigma \simeq \theta_{2\pi}$, in the corresponding formulas to 
Eqs.~(\ref{c4}) and (\ref{c5}). 
The amplitude has zero close to the threshold and 
no direct $\sigma$ Breit-Wigner peak is observed in $\pi\pi$ mass spectra. 

On the other hand, for general $\pi\pi$ production processes 
the parameters $r_i$ and $\theta_i$ are independent of those in  
$\pi\pi$ scattering, since they are concerned with different 
$S$-matrix elements.
Especially we can expect in the case of ``$\sigma$-dominance", 
$r_\sigma \gg r_{2\pi}$, the $\pi\pi$ spectra 
show steep increase from the $\pi\pi$ threshold, and the 
$\sigma$ Breit-Wigner peak is directly observed. 
This situation seems to be realized in 
$J/\psi\rightarrow\omega\pi\pi$ and $D^+\rightarrow\pi^-\pi^+\pi^+$.

Here we should note that the chiral constraint on 
$r_\sigma$ and $r_{2\pi}$ does not work 
generally in the production processes 
with large energy release to the $\pi\pi$ system.
We explain this fact in case of $\Upsilon$ decays.\cite{PLB2} Here we take 
$J/\psi\rightarrow\omega\pi\pi$ as an example.
We consider a non-resonant $\pi\pi$ amplitude of derivative-type 
${\cal F}_{\rm der}$, 
\begin{eqnarray}
{\cal F}_{\rm der} &\sim&  P_\psi \cdot p_{\pi 1} P_\psi \cdot p_{\pi 2} / M_\psi^2 , 
\label{c7}
\end{eqnarray}
where $P_\psi (p_{\pi i})$ is the momentum of $J/\psi$ 
(emitted pions).\footnote{
The equation (\ref{c7}) is obtained by the chiral symmetric effective 
Lagrangian,
${\cal L}_d=\xi_d \partial_\mu \partial_\nu \psi_\lambda \omega_\lambda 
(\partial_\mu \pi \partial_\nu \pi + \partial_\mu \sigma \partial_\nu \sigma )$.
The possible origin of this effective Lagrangian is discussed in ref.\citen{PLB2}. 
There occurs no one $\sigma$-production amplitude, cancelling the $2\pi$ amplitude,  
in this Lagrangian. 
}
This type of amplitude satisfies the Adler self-consistency condition,
${\cal F}\rightarrow 0$ when $p_{\pi 1\mu}\rightarrow 0$, and consistent with
the general chiral constraint. 
However, this zero does not appear in low energy region of actual $s$-plane.
At $\pi\pi$ threshold (where $s=4m_\pi^2$),
$p_{\pi 1\mu}= p_{\pi 2\mu}$ and  
$P_\psi \cdot p_{\pi i} / M_\psi=E_{\pi i} \sim 
(M_\psi -m_\omega )/2$ ($E_{\pi i}$ being the energy of emitted pion), 
and thus, 
\begin{eqnarray}
{\cal F}_{\rm der} &\rightarrow& (\ (M_\psi -m_\omega )/2\ )^2 \gg m_\pi^2 \ \ \ {\rm at}\ \ \ 
s\rightarrow 4m_\pi^2.
\label{c8}
\end{eqnarray}
The amplitude (\ref{c7}) is not suppressed near $\pi\pi$ threshold, and 
correspondingly there is no strong constraint between $r_\sigma e^{i\theta_\sigma}$
and $r_{2\pi}e^{i\theta_{2\pi}}$ leading to the threshold suppression.
This is quite in contrast with the situation in $\pi\pi$ scattering, 
Eq.~(\ref{c6}). 

({\it ``Universality" of ${\cal T}_{\pi\pi}$ : 
threshold behavior})\ \ \ \ 
Conventionally all the production amplitudes ${\cal F}_{\pi\pi}$,
including the $\pi\pi$ system in the final channel, 
are believed\cite{pen} to take the form 
proportional to ${\cal T}_{\pi\pi}$ as
\begin{eqnarray}
{\cal F}_{\pi\pi} &=& \alpha (s) {\cal T}_{\pi\pi};
\ \ \alpha (s):{\rm slowly\ varying \ 
                       real\ function} , 
\label{c9}
\end{eqnarray} 
where $\alpha (s)$ is supposed to be a slowly varying real function.
This implies that 
${\cal F}$ and ${\cal T}$ have the same phases and the same structures
(the common positions of poles, if they exist).
The equation (\ref{c9}) is actually applied to the analyses of various production  
processes\cite{pen,bug,Zou,Anisovich}, and it was the reason of 
overlooking $\sigma$
for almost 20 years in the 1976 through 1994 editions of Particle Data Group tables. 

The equation (\ref{c9}) is based on the belief that 
low energy $\pi\pi$ chiral dynamics
is also applicable to $\pi\pi$ production processes with small $s$,
leading to the threshold suppression of spectra, as in  Eq.~(\ref{c6}), 
in all production processes
because of the Adler zero in ${\cal T}_{\pi\pi}$. This 
is apparently inconsistent with experimental data.

So, in order to remove the Adler zero at $s=s_0$ and 
to fit the experimental spectra, one is forced to modify\cite{pen,bug} 
the form of $\alpha (s)$ by multiplying artificially 
the rapidly varying factor $1/(s-s_0)$
without any theoretical reason.

When the scattering amplitude ${\cal T}$ is unitarized by $N/D$ method,
Adler zero in $N$ leads to the factor $s-s_0$ in the imaginary part of $D$. 
Then, in ref. \citen{bug}, following Eq.~(\ref{c9}),
in addition to the (above mentioned) artificial factor 
in $\alpha (s)$, 
it is insisted that this $D$ function 
with the factor $s-s_0$ in its imaginary part
should be applied to all the $\pi\pi$ production 
amplitudes ${\cal F}$.
However, {\it such a requirement on ${\cal F}$ has no relation with
the Adler self-consistency condition which predicts the zero in total ${\cal F}$.
Moreover, that is not generally valid since in its approach only the $\pi\pi$
dynamics, that is, the final state interaction between two stable $\pi$ mesons
is considered, and the various final state interactions, as expalined in Fig. \ref{figFSI},
are not taken into account.}
In the case of $J/\psi\rightarrow\omega\pi\pi$, 
the emitted pion energy is of order $M_\psi$,
and the new dynamics in $J/\psi$ energy region, which is beyond the scope
of chiral dynamics, must also be considered.
This is overlooked\cite{pen,bug} in Eq.~(\ref{c9}).

Here I should like to stress the physical meaning of an example, Eq.~(\ref{c7}),
that Adler zero condition, even in the isolated final $2\pi$ system, does not necessarily
lead to the threshold suppression. 
This implies that the above mentioned ad hoc prescription\cite{bug}
to get rid of the undesirable zero near threshold of ${\cal F}$ 
becomes not necessary, if we take the new dynamics duely into account.

\section{Phases of Production Amplitudes}

\hspace*{-0.8cm}({\it Generalized $S$ matrix and 
phase of production amplitude})\ \ \ \ 
It is often discussed that in order to confirm the existence of a resonant 
particle,
it is necessary to observe the corresponding phase motion 
$\Delta \delta \sim 180^\circ$ 
of the amplitude like the case of $\rho$ meson. 
In the $\pi\pi\ P$ wave amplitude 
a clear phase motion $\Delta \delta \sim 180^\circ$ due to $\rho$ meson 
Breit-Wigner amplitude is observed.
However, in the case of $\sigma$ meson, 
because of the chiral cancellation mechanism (explained in \S 2), and 
because of its
large width,
the $\sigma$ Breit-Wigner phase motion $\Delta \delta \sim 180^\circ$
cannot be observed directly in the $\pi\pi$ scattering amplitude.
While, {\it in $\pi\pi$ production processes the amplitudes are the sum of 
various $S$ matrix elements
as explained in Eqs.~(\ref{c4}) and (\ref{c5}), and correspondingly
the pure $\sigma$ Breit-Wigner phase motion may be generally difficult 
to be observed.}
However, only in some exceptional cases, 
when the amplitude is dominated by $\sigma$,
the $\sigma$ phase motion may be directly observed. 

On the other hand, as was mentioned at the end of \S 3, conventionally  
it is widely believed that all the $\pi\pi$ production amplitude ${\cal F}$
have the same phase as that of $\pi\pi$ scattering amplitude ${\cal T}$.

However, {\it this belief (or Eq.~(\ref{c9})) comes from the 
incorrect application of elastic unitarity condition,
which is not applicable to production processes},
where the freedom of various strong phases, 
$\theta_\sigma$, $\theta_{b_1}$, $\theta_{f_2}$,
$\cdots$ (in Eqs.~(\ref{c4}) and (\ref{c5})) 
allowed in {\it generalized unitarity condition}, is overlooked 
in Eq.~(\ref{c9}). 

{\it Because of the effect of the above strong phases, generally ${\cal F}$ 
have different phases from ${\cal T}$.}  
${\cal F}$ have the same phase as ${\cal T}$ only in very limited cases
when the final $\pi\pi$ (or $K\pi$) systems are completely
isolated in strong interaction level. For 3-body decays such as 
$J/\psi\rightarrow \omega\pi\pi (K^*K\pi)$ and 
$D^-\rightarrow\pi^-\pi^+\pi^+(K^-\pi^+\pi^+)$,
the above condition is not satisfied.
Actually, the large strong phases in $J/\psi$ and $D$ decays are suggested
experimentally.
In order to reproduce the experimental branching ratio of 
$J/\psi\rightarrow 1^-0^-$ decays 
(that is, $J/\psi\rightarrow \omega\pi^0,\rho\pi,K^*\bar K,\cdots$,)
it is necessary to introduce a large relative strong phase\cite{Mahiko} 
$\delta_\gamma =arg\frac{a_\gamma}{a}=80.3^\circ$ 
between the effective coupling constants
of three gluon decay $a$ and of one photon decay $a_\gamma$.
A similar result is also obtained in  $J/\psi\rightarrow 0^-0^-$ decays.
A large relative phase between $I=3/2$ and $I=1/2$ amplitudes 
of $D\rightarrow K\pi$ decays is observed: 
$\delta_{3/2}(m_D)-\delta_{1/2}(m_D)=(96\pm 13)^\circ$,\cite{Dphase}
(while in $B\rightarrow D\pi ,D\rho ,D^*\pi$ decays rather small relative phases are obtained).
By considering these results we expect not small strong phases $\theta_\sigma$, $\theta_{b_1}$,
$\cdots$ coming from $\sigma \omega ,b_1\pi ,\cdots$ rescatterings 
in $J/\psi$ decays.\footnote{
It is often argued that the amplitude of $J/\psi\rightarrow\omega\pi\pi$ 
(or $D\rightarrow\pi^-\pi^+\pi^+$ ) near $\pi\pi$ threshold 
must take the same phase as the $\pi\pi$ scattering phase, 
since in this energy region the $m_{\omega\pi} (\sim M_\psi )$ is large, 
and $\pi\pi$ decouples from $\omega$ in final channel.\cite{Tuan} 
And this phase constraint is argued to come 
from $\pi\pi$ elastic unitarity condition.
However, according to the work\cite{Mahiko} $M_\psi$ is not sufficiently large 
for making $\pi\pi$ decouple from $\omega$, and the $\pi\pi$ elastic 
unitarity constraint actually does not work in the amplitude Eq.~(\ref{c4}).
} 

({\it cos $\theta$ distribution in 
$J/\psi\rightarrow\omega\pi\pi$})\ \ \ \ 
In relation to this argument
Minkowski and Ochs raise a criticism 
concerning the existence of light-mass $\sigma$-pole\cite{MO1} 
in $J/\psi\rightarrow\omega\pi\pi$:

The cos$\theta$ distribution is obtained 
in $m_{\pi\pi}$=250$\sim$750MeV~by~DM2.\cite{DM2}
They apply partial wave expansion(PWA) including $S$ and $D$ waves 
to obtain the cross section as
$
\frac{d\sigma}{d\Omega} \sim |S|^2 + 10 ( 3 {\rm cos}^2\theta -1) Re(SD^*) +{\cal O}(|D|^2),
$
($S(D)$ is the $\pi\pi$ S(D)-wave component), where 
the cos$^2\theta$ term is proportional to $Re(SD^*)$.
Then, if the $D(S)$ is dominated by $f_2(1270)\ (\ \sigma\ )$ contribution,
the angular distribution would vary with a sign change of 
the cos$^2\theta$ term (from $+$ to $-$).
Actually, contradictorily to this anticipation, 
the data do not show any sign change below 750 MeV,
and they conclude that there is no indication for
a Breit-Wigner resonance at 500 MeV.\cite{MO1}

However, (according to our preliminary analyses,\cite{sigmaG}) in this energy region 
there is almost no contribution from $f_2(1275)$, and 
actually the $l\geq 2$ partial waves mainly come from $b_1(1235)$ contribution,
$J/\psi\rightarrow \pi b_1$ and $b_1\rightarrow \omega\pi$.
In $m_{\pi\pi}\stackrel{>}{\scriptscriptstyle \sim} 500$MeV the direct $b_1$ peak 
is seen in cos $\theta$ distribution, and it includes
the large higher wave components.
This fact means the above  PWA does not work well in this energy region. 
Furthermore, each of the partial waves is expected to show 
a large phase movement (from $b_1$ pole)
in the relevant energy region $m_{\pi\pi}\sim 500$MeV, while in the above criticism
the almost constant phase of $D$ wave is assumed.  
Thus, the basic assumption is not applicable,
and their criticism is not correct. 

Recently
a method extracting the $\sigma$ phase motion from Dalitz plot data of $D$ decays
is presented in ref. \citen{brazil}, where
the interference between $f_0(980)$ (or $f_2(1275)$) 
and the remaining $S$-wave component is used to observe the $\sigma$ phase motion.
Similarly, in the relevant $J/\psi$ decays, by using the Dalitz plot data, 
it may be possible to observe the $\sigma$ phase motion, where
the interference between $b_1(1235)$ Breit-Wigner amplitude 
and $\pi\pi$ $S$-wave component is used.
The direct $b_1$ peak appears in $m_{\pi\pi}\stackrel{>}{\scriptscriptstyle \sim} 500$MeV region
of Dalitz plot, and in this energy region 
the $S$-wave phase motion is expected to be determined 
with good accuracy.\footnote{In case 
$(m_\sigma ,\Gamma_\sigma )=(500,350)$MeV the $\sigma$
Breit-Wigner amplitude gives the phase difference 
$\Delta\delta =83^\circ (67^\circ)$
between $m_{\pi\pi}$=450$\sim$850MeV(500$\sim$850MeV)
which is somewhat larger than the corresponding $\pi\pi$ scattering phase difference
$\Delta\delta \simeq 63^\circ (55^\circ)$.}

\begin{wrapfigure}{r}{8cm}
\vspace*{-1em}
  \epsfysize=4.cm
  \centerline{\epsffile{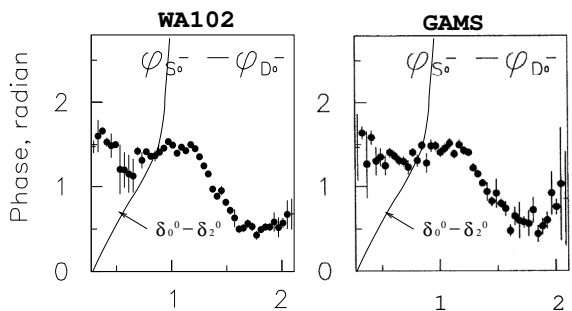}}
\vspace*{-0.5em}
  \caption{The relative phase $\phi_{S_0^-}-\phi_{D_0^-}$
between $S_0^-$ and $D_0^-$ in $pp\rightarrow pp \pi^0\pi^0$ in
(a) WA102\cite{WA102} and (b) GAMS.\cite{GAMS}
$\phi_{S_0^-}-\phi_{D_0^-}$ is different from the scattering 
$\delta_0^0-\delta_2^0$ shown by solid line. 
 }
  \label{figCentral}
\end{wrapfigure}
({\it $pp$ central collision, 
$pp\rightarrow pp \pi^0\pi^0$})\ \ \ \ 
A partial wave analyses of the $\pi^0\pi^0$ system produced centrally 
in $pp$ collisions at 450 GeV$/$c are done by WA102\cite{WA102} and GAMS\cite{GAMS}.
A large peak structure around 500 MeV, observed in both $S$ and $D$ waves,
is explained in ref. \citen{MO1} by the one pion exchange (OPE).

It should be noted that the relative phase $\phi_{S_0^-}-\phi_{D_0^-}$
between $S_0^-$ and $D_0^-$ amplitudes\footnote{ 
The relative phase between $S_0^-$ and $D_1^-$ in the relevant energy region
is different from that between $S_0^-$ and $D_0^-$.
The $D_1^\pm$ components explain the $\phi$ distribution,
which suggests the pomerons have vector components. }
in Fig. \ref{figCentral}
is apparently different from the corresponding scattering phase $\delta_0^0-\delta_2^0$,
(shown by solid line in the figure)
where $\delta_0^0$ and $\delta_2^0$ are isosinglet  
$S$ and $D$ wave $\pi\pi$ phase shifts, respectively.  
(As shown in Fig.\ref{figpipi}, $\delta_0^0$ gradually increases from 0 to 90$^\circ$ 
in $m_{\pi\pi}=2m_\pi$ through $\sim 900$ MeV,
and $\delta_2^0$ dominated by $f_2(1275)$ is almost 0 in this energy region.)
Thus, {\it it is experimentally confirmed that the phase of $\pi\pi$ 
production amplitude 
${\cal F}$ is different from $\pi\pi$ scattering amplitude ${\cal T}$} 
in this process.

\begin{wrapfigure}{r}{6cm}
\vspace{-1em}
  \epsfysize=3.5cm
  \centerline{\epsffile{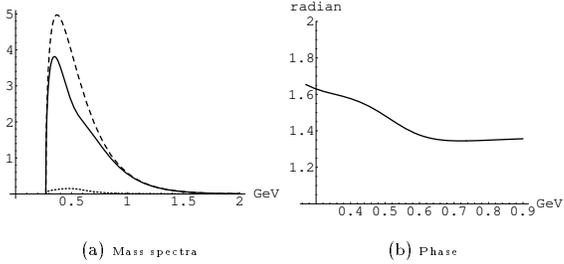}}
\vspace{-0.5em}
  \caption{(a) Mass spectra and (b) phase below 0.9 GeV of the amplitude 
            ${\cal F}$ shown by solid line. 
            In (a) dashed(dotted) line represents the background ($\sigma$) 
            contribution with arbitrary scale. 
            The values of parameters, 
            $b=0.5,c=5,a=100,\theta_{BG}=1.5,r_\sigma =0.05 ,
            \theta_\sigma =\pi$ (in appropriate order of GeV unit), 
            are selected by inspection.
 }
  \label{figPPphase}
\end{wrapfigure}
As shown in Fig.\ref{figCentral}, in both experiments, the $\phi_{S_0^-}-\phi_{D_0^-}$ 
is not constant, and 
shows some structure around $0.5 \sim 0.6$ GeV, which seems to suggest 
the $\sigma$ contribution interfering with the background which may come from OPE.
In VMW method the ${\cal F}$ is given by\\
${\cal F} = r_{BG}(s) e^{i\theta_{BG}} + r_\sigma e^{i\theta_\sigma}
\frac{m_\sigma\Gamma_\sigma (s)}{m_\sigma^2-s-im_\sigma \Gamma_\sigma (s)} ,
$
where the 1st(2nd) term represents the background($\sigma$ BW) amplitude.
An example of the mass spectra, $|{\cal F}|^2/\rho (s)$ 
($\rho (s)$ being $\pi\pi$ state density $p_1(s)/(8\pi\sqrt{s})$), and the phase $Arg {\cal F}$
is shown  in Fig.~\ref{figPPphase}.
Here we use $r_{BG}^2(s)={a(\sqrt s - 2m_\pi^0)^b {\rm exp} (-c\sqrt s -ds)} \times \rho (s)$,
which is used in ref.\citen{WA102,GAMS}, and 
the parameters are selected by inspection of experimental mass and phase distributions.
Experimental data in this process seem to be consistent with 
the existence of $\sigma$ meson.

({\it $D^+\rightarrow K^-\pi^+\mu^+\nu$ and $D^+\rightarrow K^-\pi^+\pi^+$ })\ \ \ \ 
The $K^-\pi^+$ spectra of $D^+\rightarrow K^-\pi^+\mu^+\nu$ by 
FOCUS\cite{FOCUS}
is dominated by $P$ wave from $\bar K^{*0}$ interfering with a small
$S$ wave. Through the angular analysis this $S$ wave component
has almost constant phase $\delta =\frac{\pi}{4}$ 
in mass region of $\kappa$ meson,
$m_{K\pi}=0.8\sim 1.0$GeV. 
This $\delta$ is suggested, by Minkowski and Ochs\cite{MO1},
to be the same as the $K\pi$ scattering phase shift\cite{LASS} by LASS in this mass region.
Based on this result, they critisize the analysis of $D^+\rightarrow K^-\pi^+\pi^+$ 
by E791\cite{E791kappa}, where, as explained in \S2, the $\kappa$ Breit Wigner amplitude 
is applied and it has large phase motion in this mass region. 
They stated 
``Such a result (of E791) appears to contradict the above FOCUS result.''

However, this criticism is again premature because the effect of strong phases
allowed in generalized unitarity condition is overlooked.
In $D^+\rightarrow K^-\pi^+\mu^+\nu$ the final $K^-\pi^+$ is isolated 
in strong interaction level.
Thus, as we explained in the first part of this section, 
the amplitude has the same phase as
$K\pi$ scattering amplitude due to Watson theorem.
However,  $D^+\rightarrow K^-\pi^+\pi^+$ is a three body decay of heavy meson
and $K^-\pi^+$ is not isolated in final channel, and the rescattering effects
from various elements of generalized $S$ matrix, such as, $_{out}\langle\bar\kappa^{0}\pi^+|J/\psi\rangle_{in}$,
$_{out}\langle\bar K^{*0}\pi^+|J/\psi\rangle_{in}$, $_{out}\langle\bar K_2^{*0}\pi^+|J/\psi\rangle_{in}$,
$\cdots$, must be taken into account. 
In 2-body decays of $D$, $J/\psi$ ($B$) mesons, the large (small) strong phase is 
suggested.\cite{Mahiko} Thus, the phases of the above matrix elements
are considered to be not small.
Thus, 
total amplitude of $D^+\rightarrow K^-\pi^+\pi^+$ 
has generally different phase from that of 
$D^+\rightarrow K^-\pi^+\mu^+\nu$.

Finally we should add a comment on $D^+\rightarrow K^-\pi^+\mu^+\nu$ 
in relation to $K_{l4}$ decay $K^+\rightarrow \pi^+\pi^- e^+\nu$.
The latter process is analyzed by Shabalin\cite{Shabalin} by using SU(3) linear $\sigma$ model,
where the effect of direct $\sigma$ production, 
$K^+\rightarrow \sigma e^+\nu $ 
(successively $\sigma\rightarrow \pi^+\pi^-$), explains 
the large width obtained experimentally 
(which is twice as large as the prediction by soft pion limit).
At the same time this amplitude has the same phase 
as the $\pi\pi$ scattering phase shift. 
{\it The $\sigma$ Breit Wigner phase motion is not observed 
due to Watson theorem, 
but its large decay width suggests the $\sigma$ production} 
in this process. As can be seen in this example,
the $\kappa$ phase motion is not observed in  
$D^+\rightarrow K^-\pi^+\mu^+\nu$, but this fact
does not mean no $\kappa$-existence. 
The analysis of $D^+\rightarrow K^-\pi^+\pi^+$ by E791 does not contradict 
with FOCUS result, and strongly suggests the $\kappa$(800) existence.

\section{Concluding Remarks}

The masses and widths of $\sigma$ and $\kappa$ mesons 
quoted in the text are summarized in Table \ref{tabProperty}.  
The peak structures of $\sigma$ and $\kappa$ meson productions
are directly observed in these $\pi\pi$ and $K\pi$ production processes.
It is remarkable that a clear peak structure from $\kappa$ production 
is observed in $K\pi$ mass spectra of 
$J/\psi\rightarrow K^{*0}K^-\pi^+$ by BES,\cite{WNKomada}
which is added in the table.

\begin{wraptable}{r}{8cm}
\begin{tabular}{l|c|c}
Processes & $m_\sigma$(MeV) & $\Gamma_\sigma$(MeV) \\
\hline
$\pi\pi\rightarrow\pi\pi$ & 535$\sim$675 & $385\pm 70$\\
$J/\psi\rightarrow\omega\pi\pi$(DM2) & 482 & 325\\
$J/\psi\rightarrow\omega\pi\pi$(BES) & 
             $390\stackrel{+60}{\scriptstyle -36}$
               &  $282\stackrel{+77}{\scriptstyle -50}$\\
$\Upsilon (mS) \rightarrow \Upsilon (nS)\pi\pi$ & $526\stackrel{+48}{\scriptscriptstyle -37}$
               &  $301\stackrel{+145}{\scriptstyle -100}$\\
$p\bar p \rightarrow 3\pi^0$ & $540\stackrel{+36}{\scriptstyle -29}$
               &  $385\stackrel{+64}{\scriptstyle -80}$\\
$D^+ \rightarrow \pi^-\pi^+\pi^+$ & $478\stackrel{+24}{\scriptstyle -23}\pm 17$
               &  $324\stackrel{+42}{\scriptstyle -40}\pm 21$\\
$\tau^- \rightarrow \pi^-\pi^0\pi^0\nu_\tau$ & 555 & 540\\
\hline\hline
Processes & $m_\kappa$(MeV) & $\Gamma_\kappa$(MeV) \\
\hline
$K\pi \rightarrow K\pi$ & $905\stackrel{+65}{\scriptstyle -30}$
               &  $545\stackrel{+235}{\scriptstyle -110}$\\
$D^+ \rightarrow K^-\pi^+\pi^+$ & $797\pm 19 \pm 42$
               &  $410 \pm 43 \pm 85$\\
$J/\psi\rightarrow K^{*0} K^-\pi^+$ & $\sim 800$ & $\sim 300$ \\ 
\end{tabular}
\caption{The mass and width of $\sigma$ and $\kappa$ mesons.}
\label{tabProperty}
\end{wraptable}
The peak structures mentioned above are fitted well by the Breit-Wigner amplitudes of $\sigma$ and $\kappa$ 
following VMW method, independently from the $\pi\pi$ and $K\pi$ scattering. 
The criticisms\cite{MO1,bug} on this method
come from the confusion of the 3-body production processes
with 2-body scattering process, and are not correct.

The recent belief that phases of all the $\pi\pi$ production amplitude 
${\cal F}$ is the same as that of ${\cal T}$ also 
comes from the same kind of confusion, and thus, is not correct.   
The 3- (or multi-) body production processes including $\pi\pi$ or $K\pi$ system in the final channels
are considered as superpositions of various two- (or more) bodies processes,
which correspond to the different elements of general $S$-matrix.   
Thus, the total amplitude generally has different phase from that of the $\pi\pi$
and $K\pi$ scatteirng amplitudes.

The observed peak structures are considered as the strong evidences of $\sigma$ and $\kappa$ existence.
Presently their existence seems to be established 
with the property $(m_\sigma ,\Gamma_\sigma )=(\sim 500, \sim 300)$MeV and
$(m_\kappa ,\Gamma_\kappa )=(\sim 800, \sim 400)$MeV, respectively. 
  

{\it 
The author would like to express his gratitudes 
to Prof. S. F. Tuan for useful information.
He also thanks to Prof. M. Oka, Prof. S. Ishida , Prof. K. Takamatsu and 
Prof. T. Tsuru for discussions, and  
to Dr. T. Ishida for making figures.
}

\end{document}